\documentclass{jfm}

\usepackage{graphicx}
\usepackage{natbib}

\usepackage{amsmath}
\usepackage{subcaption}
\usepackage{bm}

\ifCUPmtlplainloaded \else
  \checkfont{eurm10}
  \iffontfound
    \IfFileExists{upmath.sty}
      {\typeout{^^JFound AMS Euler Roman fonts on the system,
                   using the 'upmath' package.^^J}%
       \usepackage{upmath}}
      {\typeout{^^JFound AMS Euler Roman fonts on the system, but you
                   dont seem to have the}%
       \typeout{'upmath' package installed. JFM.cls can take advantage
                 of these fonts,^^Jif you use 'upmath' package.^^J}%
      }
  \else
  \fi
\fi

\ifCUPmtlplainloaded \else
  \checkfont{msam10}
  \iffontfound
    \IfFileExists{amssymb.sty}
      {\typeout{^^JFound AMS Symbol fonts on the system, using the
                'amssymb' package.^^J}%
       \usepackage{amssymb}%
         \let\leq=\leqslant
         
      }{}
  \fi
\fi

\ifCUPmtlplainloaded \else
  \IfFileExists{amsbsy.sty}
    {\typeout{^^JFound the 'amsbsy' package on the system, using it.^^J}%
     \usepackage{amsbsy}}
    {\providecommand\boldsymbol[1]{\mbox{\boldmath $##1$}}}
\fi

\providecommand\bcdot{\boldsymbol{\cdot}}

\newsavebox{\astrutbox}
\sbox{\astrutbox}{\rule[-5pt]{0pt}{20pt}}

\usepackage{color}
\definecolor{light-gray}{gray}{0.5}
\definecolor{blue}{rgb}{0.0,0.0,1.0}
\definecolor{green}{rgb}{0.0,0.5,0.0}
\definecolor{red}{rgb}{1.0,0.0,0.0}
\definecolor{cyan}{rgb}{0.0,0.75,0.75}
\definecolor{magenta}{rgb}{0.75,0.0,0.75}
\definecolor{yellow}{rgb}{0.75,0.75,0.0}

\title[Breaking Kelvin]{{Breaking Kelvin: Circulation conservation and vortex breakup in MHD at low Magnetic Prandtl Number}}
\author{ D.G.~Dritschel${\ddag}$, P.H. Diamond \&  S.M. Tobias${\dag}$$^{\ast}$\thanks{$^\ast$Corresponding author. Email: smt@maths.leeds.ac.uk}}

\shorttitle{Breaking Kelvin} 
\shortauthor{ D.G.~Dritschel, P.H. Diamond \& S.~M.~Tobias} 

\affiliation{Department of Applied Mathematics, University of Leeds, Leeds LS2 9JT, UK}

\pubyear{2010}
\volume{650}
\pagerange{119--126}
\date{?; revised ?; accepted ?. - To be entered by editorial office}
\begin{document}

\maketitle

\begin{abstract}
In this paper we examine the role of weak magnetic fields in breaking
Kelvin's circulation theorem and in vortex breakup in two-dimensional
magnetohydrodynamics for the physically important case of a low magnetic
Prandtl number (low $Pm$) fluid. We consider three canonical inviscid solutions
for the purely hydrodynamical problem, namely a Gaussian vortex, a
circular vortex patch and an elliptical vortex patch. We examine how
magnetic fields lead to an initial loss of circulation $\Gamma$ and
attempt to derive scaling laws for the loss of circulation as a
function of field strength and diffusion as measured by two
non-dimensional parameters. We show that for all cases the loss of
circulation depends on the integrated effects
of the Lorentz force, with the patch cases leading to significantly
greater circulation loss. For the case of the elliptical vortex the
loss of circulation depends on the total area swept out by the
rotating vortex and so this leads to more efficient circulation loss
than for a circular vortex.
\end{abstract}
\newcommand{\bu}{{\boldsymbol u}}
\newcommand{\bj}{{\boldsymbol j}}
\newcommand{\bB}{{\boldsymbol B}}
\newcommand{\bF}{{\boldsymbol F}}
\newcommand{\bk}{{\boldsymbol k}}
\newcommand{\htil}{{\tilde h}}
\newcommand{\bOmega}{{\boldsymbol \Omega}}
\newcommand{\bomega}{{\boldsymbol \omega}}
\newcommand{\bnabla}{{\boldsymbol \nabla}}
\newcommand{\oeps}{{\cal O} (\epsilon)}
\newcommand{\oepssq}{{\cal O} (\epsilon^2)}
\newcommand{\pre}{Physical Review E}
\newcommand{\apj}{Astrophysical Journal}
\newcommand{\apjl}{Astrophysical Journal Letters}
\newcommand{\nat}{Nature}
\newcommand{\icarus}{Icarus}
\newcommand{\solphys}{Solar Physics}

\providecommand\bcdot{\boldsymbol{\cdot}}
\newcommand{\bx}{{\boldsymbol x}}
\newcommand{\bex}{\hat{\boldsymbol e}_x}
\newcommand{\bey}{\hat{\boldsymbol e}_y}
\newcommand{\gperp}{{\boldsymbol \nabla}^\perp}
\newcommand{\lap}{\nabla^2}
\newcommand{\beq}{\begin{equation}}
\newcommand{\eeq}{\end{equation}}


\section{Introduction}

The interaction of magnetic fields and (often turbulent) fluid flows is a fundamental problem of geophysical and astrophysical fluid dynamics. Its importance lies in the fact that this interaction often controls the dynamics of geophysical and astrophysical objects, with amplification of the field via stretching by turbulent flows leading to dynamo action \citep[see e.g.][]{moffatt78} and the back-reaction of the magnetic field on the turbulent flows via the Lorentz force. This  leads to, for example, the suppression and excitation of instabilities \citep{rkh_2013} and the modification of the mixing and transport properties of the flow \citep{moffatt_1983,GD_1994}. These interactions may occur in models of self-consistent dynamo action \citep[see e.g.][]{ds_2007}, magnetoconvection \citep{wp_2014} or stably stratified magnetic layers \citep{t_2010}.

In this paper, we shall be concerned with the interaction of magnetic fields in fluid layers that are stably stratified. Such layers are common in geophysics and astrophysics, with notable examples being the solar tachocline \citep[see e.g.][and the references therein]{hrw_2012}, the stably stratified layer at the core-mantle boundary \citep{gd_2013,buff_2014}, and possible stable layers in gas giants and exoplanets \citep{s_2003,cw_2008}. It is well known that stable stratification leads to strong anisotropy in flows with vertical (i.e. aligned with gravity) motions inhibited relative to horizontal ones. Because of this anisotropy, the dynamics in such layers is often modelled using reduced models of varying complexity; with thin-layer models, shallow water models and two-dimensional models all providing useful insights into the dynamics of such layers. Recently the more sophisticated of these models have been extended to MHD \citep[see e.g.][]{g_2000,mg_2004}, though much progress can still be made utilising basic two-dimensional models; it is this approach we use here, though we discuss generalisations to such models in the conclusions.

It is well known that rotating, stably stratified fluids exhibit correlations that can lead to the generation of large-scale flows. There are many explanations for such behaviour, but it is often argued that the existence of conservation laws and an inversion procedure plays a key role in this emergence of large-scale behaviour \citep{DM:2008}. For example in two-dimensional dynamics vorticity (and in quasi-two-dimensional dynamics potential vorticity) are important conserved quantities.  \citet{TDH_2007} have demonstrated that the presence of even a weak magnetic field may lead to the suppression of the correlations that lead to the formation of large-scale flows. A central question then is what role the magnetic field may play in the modification of the Lagrangian conserved quantities of the flow, via the action of the Lorentz force \citep{KSD_2008}. Furthermore, as described below,  it has been demonstrated that magnetic fields may inhibit shear flow instabilities and lead to the disruption of coherent structures such as vortices.

One key issue (often conveniently overlooked) for the modelling of astrophysical stable layers  is that both the ionised plasma of stellar interiors and the liquid iron of the Earth's fluid outer core are fluids with
extremely low magnetic Prandtl number $Pm = Rm/Re = \nu/\eta$, where $\nu$ is the viscosity and $\eta$ is the magnetic diffusivity of the fluid. This has the consequence that the velocity or vorticity field dissipates at spatial scales much smaller than the magnetic field (which is itself at small scales owing to high values of $Rm$) \citep{tcb:2013}. Thus the description of low $Pm$ dynamics is a computationally difficult problem, and virtually all numerical investigations have been carried out with $Pm \sim {\cal O}(1)$.

Here we briefly mention some previous investigations of the interaction of weak magnetic fields with two dimensional flows. Note this body of work is distinct from that examining the dynamics of MHD turbulence with a strong guide field (where quasi two-dimensionality occurs in the plane perpendicular to the magnetic field; for a review see \citet{tcb:2013}). Weak magnetic fields are known to have a significant effect on two dimensional turbulence. Most striking is that the presence of a magnetic field can alter the direction of the spectral transfer of two dimensional turbulence, turning inverse cascades into forward cascades \citep[see e.g.][]{Pouquet_1978,sesh_etal_2014, Banpan_2014,Seshalex_2016}. The inverse cascade in two dimensional hydrodynamic turbulence is often ascribed to the global conservation of enstrophy and energy in the dissipationless regime; in the presence of magnetic field the global conservation of enstrophy is broken. 

It has also been shown that magnetic fields can have strong effects on the linear and nonlinear evolution of shear flows in two dimensions \citep{fjrg_1996, jgrf_1997, bk_2002, phzh_2008}. Here magnetic fields can interact with the vortices that arise as a result of Kelvin-Helmholtz instability and also prevent roll-up occurring in the first place.

Perhaps the most  subtle and interesting effect is that weak magnetic field can inhibit the transport (turbulent diffusion) of scalar fields as described for example by \citet{CV_1991,GD_1994,Cattaneo_1994,kht_2016}. Mixing and transport is often associated with the Lagrangian properties of fluid flows and so it is often argued that in order for a weak magnetic field to have an effect it must be subtly altering these properties, for example by breaking the material invariants of the flow. It is this effect that we shall investigate here.

All the calculations described above are performed with the magnetic Prandtl number of order unity; in this (unphysical) regime  vortex filaments have the  same spatial scale as current filaments, in contrast with the geophysical and astrophysical case where the vorticity dissipates on scales much smaller than the magnetic field. Moreover the presence of a  finite amount of viscosity ensures that there are no conservation laws even in the hydrodynamic case --- therefore it is uninteresting to examine the role of weak magnetic field in modifying conservation laws. Finally in this case (with finite viscosity) it is very difficult to find exact hydrodynamic states to perturb with a magnetic field, so that little analysis is possible;  at best plausible scaling arguments are possible.

In this paper we utilise a numerical scheme introduced by \citet{dt_2012}. This scheme, which was compared with pseudo-spectral methods for the problem of decaying MHD turbulence, allows the 
 integration of MHD flows at low $Pm$. 
Our philosophy is to investigate the role of magnetic fields in modifying the behaviour of simple paradigm flows (Gaussian vortices and circular and elliptical vortex patches). This interaction of coherent structures with magnetic fields has been much studied in the context of flux expulsion \citep{weiss_1966, mk_1983, bbg_2001,gmt_2016}. This problem has the advantage that much can be done analytically \citep{gmt_2016}. Of particular interest is that the numerical scheme we utilise (which involves evolving contours advected by the flow) is perfectly suited to evaluating the role of magnetic field in modifying the evolution of (hydrodynamically) materially conserved quantities such as circulation; as far as we are aware this is the first evaluation of this effect in the literature.

The rest of this paper is organised as follows. In section 
2 we give the mathematical framework of the model, including a description of the equations, scalings and numerical method. In section 3 we describe the results of asymptotic and numerical analysis of the interaction of magnetic fields with Gaussian vortices, circular and elliptical vortex patches, paying particular close attention to the breaking of conservation laws. We conclude in section 4 with a description of the implications of our results and possible future investigations.

\section{Mathematical framework}
\subsection{Governing equations}

We consider the two-dimensional incompressible MHD equations
in a doubly-periodic domain at
essentially zero Prandtl number (negligible viscosity), as in
\citet{dt_2012}.  The magnetic field $\bB$ is divided into a steady mean
component $B_0$ in the $x$ direction and a remaining variable part
$-\gperp{A}$ represented by a magnetic potential $A(x,y,t)$, where
$\gperp{A}=(-A_y,A_x)$ in two dimensions and subscripts $x$ and
$y$ denote partial derivatives.  That is, we take
\beq
\label{magfield}
\bB=B_0\bex-\gperp{A}.
\eeq
As the flow $\bu$ is incompressible, it is convenient to introduce a
streamfunction $\psi(x,y,t)$ in terms of which $\bu=\gperp\psi$.
Then, we can write the 2D MHD equations in terms of the vorticity
$\omega=\gperp\bcdot\bu=v_x-u_y=\lap\psi$ and 
magnetic potential $A$ as follows:
\begin{align}
\label{voreq}
\omega_t+J(\psi,\omega) &=B_0 j_x-J(A,j), \\
\label{poteq}
A_t+J(\psi,A) &=B_0\psi_x-\eta j,
\end{align}
where $j=\gperp\bcdot\bB=-\lap{A}$ is the (vertical component of the)
current density (for unit permeability $\mu_0$), $\eta$ is the
magnetic diffusivity, and $J(a,b)=a_x b_y - a_y b_x$ is the Jacobian
operator.  The terms on the r.h.s.\ of (\ref{voreq}) come from taking
the curl of the Lorentz force per unit mass.  In the absence of a
magnetic field, the vorticity is materially conserved.  Likewise, when
$\eta=0$, the total scalar potential $A+B_0 y$ is materially
conserved.  However, $\eta>0$ appears to be necessary for regularity
of the equations, though this remains unproven \citep{cao_2017}.

\subsection{Scalings and parameters}
In the results below, we consider an initial state consisting of a
vortex of mean radius $R$ and characteristic vorticity $\omega_0$
(chosen so that the maximum velocity $U_0=\tfrac12\omega_0 R$).
The initial magnetic potential $A=0$.  We additionally prescribe the
parameters $\eta$ and $B_0$.  By choosing $R$ to be a characteristic
length, and $\tfrac12\omega_0$ to be a characteristic frequency, we can
form two dimensionless parameters from $\eta$ and $B_0$, 
\beq
\label{nondim}
\delta=\frac{\sqrt{\eta/\omega_0}}{R} \quad\textrm{and}\quad
\gamma=\frac{B_0}{U_0\delta}\,.
\eeq
The parameter $\delta$ may be regarded as the ratio of the diffusive
length scale $\ell_\eta=\sqrt{\eta/\omega_0}$ to the mean
vortex radius $R$.  Notably, the familiar magnetic Reynolds number $R_m=U_0
R/\eta=\tfrac12 \delta^{-2}$, so $\delta = (2 R_m)^{-1/2}$.
The second parameter $\gamma = (2 R_m)^{1/2} B_0/U_0$ measures
the ratio of the magnetic field --- after it has been fully
intensified by sharpened gradients in $A+B_0 y$ --- to the maximum
initial velocity $U_0$.  Hence $\gamma = 1$ denotes the field strength that would in the absence of other, more subtle interactions, bring the small-scale field into equipartition with the flow. Note, diffusion limits the gradients in
$A+B_0 y$ to
${\mathcal{O}}(B_0 R/\ell_\eta)={\mathcal{O}}(B_0/\delta)$.
So, even a weak initial field can have a strong effect if $\eta$ is
sufficiently small.

\subsection{Conservation laws}
The presence of a magnetic field breaks many hydrodynamical conservation
laws, the most important of which is conservation of circulation
\beq
\label{circ}
\Gamma=\oint_{{\mathcal{C}}} \bu\bcdot d\bx
\eeq
where ${\mathcal{C}}$ is any material contour, and $\bx$ lies on
${\mathcal{C}}$.  In an inviscid fluid, considered here, $\Gamma$ remains
constant in the absence of a magnetic field,
a fundamental result known as Kelvin's circulation theorem.  This result
is a direct consequence of material (pointwise) conservation of vorticity.

A magnetic field breaks this conservation.  The tension created by
twisting field lines tends to retard the vortex rotation, reducing
circulation in magnitude (at least initially).  The finite magnetic
diffusivity limits the build up of this tension, leading typically to
the expulsion of the field from the vortex core \citep{weiss_1966}.
In this way, only a fraction of the vortex circulation may be removed
by the action of the field. This process is subtle, indeed even in the kinematic regime it becomes apparent that it is the integrated effects of diffusion that
control the scaling of the maximum field strength \citep{weiss_1966,moffatt78}.

An expression for the rate of change of circulation can be obtained by
taking a time derivative of (\ref{circ}) using (\ref{voreq}).  One finds
\beq
\label{dcirc}
\frac{d\Gamma}{dt}=\oint_{{\mathcal{C}}} j\bnabla(A+B_0 y)\bcdot d\bx
                  =\oint_{{\mathcal{C}}} j\frac{d(A+B_0 y)}{ds}ds\,,
\eeq
where $s$ is any parametrisation of ${\mathcal{C}}$.  
Therefore, $j$ and the tangential derivative of $A+B_0 y$ along
${\mathcal{C}}$ must correlate in order to change the circulation. 

\subsection{Numerical Method}

We employ the `Combined Lagrangian Advection Method' \citep[`CLAM',
  see][]{df_2010} used in a previous work investigating 2D MHD
turbulence at low Prandtl number \citep{dt_2012}.  The method uses
material contours to represent part of the vorticity field alongside
two auxiliary gridded vorticity fields used to incorporate vorticity
forcing.  The magnetic potential $A$ is also represented on a grid,
and the pseudo-spectral method is used to calculate accurately
derivatives and invert Laplace's operator on a square doubly-periodic
domain of side length $2\pi$.  Nonlinear terms are de-aliased by
applying a circular filter in spectral space to all fields before they
are multiplied together on the grid.  The filter removes all
wavenumbers whose magnitude exceeds $2k_{\mathsf{max}}/3$, where
$k_{\mathsf{max}}$ is the maximum wavenumber in $x$ and $y$.  This is
less aggressive than the standard `2/3 rule', but is sufficient to
remove aliasing errors.  Finally, a weak $\nabla^6$ hyperdiffusion is
applied to the small-amplitude, residual vorticity field with a
damping rate of $2\omega_{\mathsf{rms}}$ at the wavenumber
$k=k_{\mathsf{max}}$.

The vorticity field is primarily represented by contours, with a
contour interval chosen to be equal to the initial range of vorticity
divided by $80$.  This is twice as fine as recommended
in \cite{df_2010} to provide higher accuracy.  Each contour is
represented by a variable number of nodes which are frequently
redistributed to maintain resolution.  This occurs each time the
`twist' $\tau$ exceeds $2.5$, where
\beq
\label{twist}
\tau=\int_{t_0}^{t} |\omega|_{\mathsf{max}}(t)dt
\eeq
and $t_0$ is the last time contour nodes were redistributed (or the
initial time).  At these times, `contour surgery' is also performed to
regularise the contours, i.e.\ to remove small filaments and to
reconnect contours of the same level which are sufficiently close.

Every $20\tau$ time units, the contours are converted to an ultra-fine
grid having dimensions $16$ times larger in each direction than the
basic `inversion' grid, and combined with the residual vorticity field
(interpolated to the ultra-fine grid) to form a gridded vorticity
field fine enough to resolve the scale of contour surgery.  This field
is then recontoured to form a new set of contours to be used for the
next $20\tau$ time units.  Recontouring acts like contour surgery but
also largely prevents contour crossing errors arising from node
redistribution.  Further details of the numerical method can be found
in \cite{df_2010} and references therein. This scheme has been demonstrated
to give extremely accurate representations of low $Pm$ dynamics;
comparable with
spectral schemes at significantly higher resolutions \citep{dt_2012}.

\section{Results}

In this section, we illustrate the flow evolution in a few
representative examples and quantify key diagnostics.  In particular,
we examine the time evolution of the energy components, mean square
vorticity and current density, and the circulation.  We determine an
appropriate scaling theory to estimate the amount of circulation
removed by the presence of a magnetic field, and contrast various
initial flow configurations: a Gaussian vortex, a Rankine vortex
(or circular vortex patch) and an elliptical vortex patch.

\subsection{Parameter settings}
In all results presented, without loss of generality we take the
characteristic vorticity within the vortex $\omega_0=4\pi$, corresponding to
a unit rotation period, and a mean vortex radius $R=5\pi/32\approx 0.49087$,
which is sufficiently small compared with the domain half width ($\pi$)
to have only a minor effect on the vortex evolution.

We consider basic `inversion' grid resolutions ranging from $128^2$ to
$1024^2$.  This grid is the one used to evolve $A$ and the gridded
vorticity fields, as well as to carry out all spectral operations,
including determining the velocity field $\bu$ from the vorticity
field.  The latter is found from the contours (after a contour to grid
conversion) and a portion of the two other gridded vorticity fields,
as described in \cite{df_2010}.  Note: the effective resolution of a
CLAM simulation is $16$ times greater in each direction, as
demonstrated in direct comparisons with a standard pseudo-spectral
method \citep{dt_2012}.

On a grid having a resolution of $n_g^2$, the magnetic diffusivity
$\eta$ is chosen as $\eta=\omega_0(\Delta x)^2$ where $\Delta
x=2\pi/n_g$ is the (basic) grid scale.  This is sufficient to resolve
the magnetic diffusion length $\ell_\eta$, as judged by the downturn
in the current density power spectrum at large wavenumber.  Note that
$\ell_\eta=\sqrt{\eta/\omega_0}=\Delta x$ with these choices of
parameters.

The two parameters we vary are $\delta=\ell_\eta/R$ and
$\gamma=B_0/(U_0\delta)$.  The former is implicitly set by the grid
resolution, and here we have $\delta=12.8/n_g$.  The second
parameter $\gamma$ is used to
determine the initial field strength $B_0$.  Given
$U_0=\tfrac12\omega_0 R=10\pi^2/32\approx 3.084$, we choose a value of
$\gamma$ and determine $B_0$ from $B_0=\gamma\delta U_0$.  For
example, for $n_g=512$, we find $\delta=0.025$, and if further
$\gamma=1$, we find $B_0\approx 0.077106$.

At the initial time $t=0$, we start with $A=0$ and one of three
vorticity distributions: (1) a Gaussian vortex with
$\omega(\bx,0)=\omega_{\mathsf{max}}e^{-r^2/2R^2}$,
where $r=|\bx|$ and
$\omega_{\mathsf{max}}=\omega_0/(2(1-e^{-1/2}))\approx1.27\omega_0$
is chosen so that the maximum velocity $U_0=\tfrac12\omega_0 R$
occurs at $r=R$; (2) a Rankine (circular) vortex with 
$\omega(\bx,0)=\omega_0$ for $r<R$ and zero otherwise; and
(3) an elliptical vortex patch having uniform vorticity $\omega_0$
within the ellipse $x^2/a^2+y^2/b^2=1$ and zero otherwise, where
$ab=R^2$ and $b/a=\lambda$, the prescribed vortex aspect ratio.
Actually, the requirement that the mean vorticity be zero within
the doubly-periodic domain implies that there is a uniform negative
compensating vorticity spread throughout the domain.  The background
vorticity is approximately $-0.024\omega_0$ for the Gaussian vortex and
$-0.019\omega_0$ for both the circular and elliptical vortex.

\subsection{Qualitative description of the flow evolution}

\begin{figure}
\begin{center}
\includegraphics[width = \textwidth]{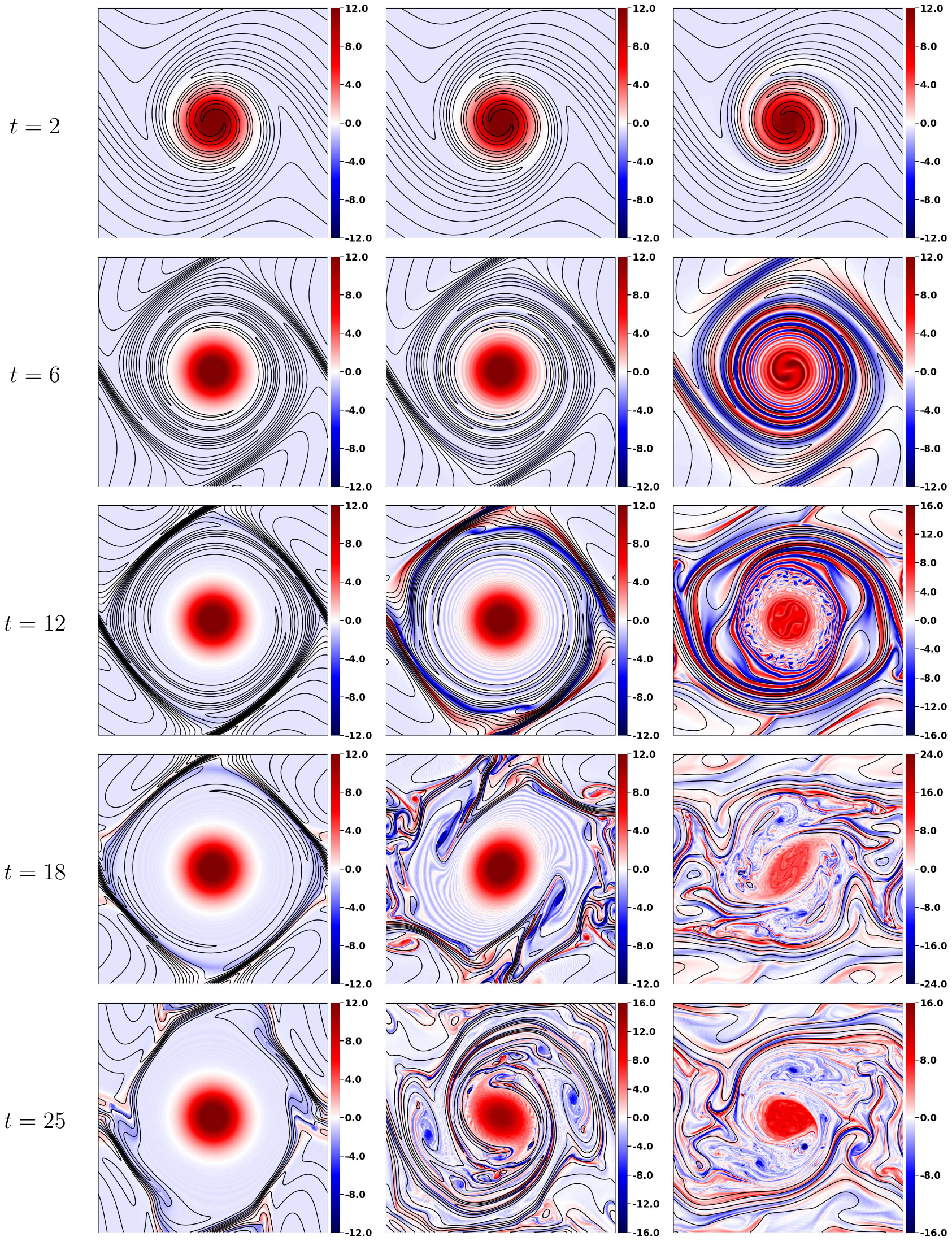}
\end{center}
\caption{Vorticity $\omega(\bx,t)$ and magnetic field lines at the
  times indicated (top to bottom) for an initially Gaussian vortex and
  for $\gamma=0.125,\,0.5$ and $2$ (left to right).  The vorticity
  colourbar is indicated next to each image.  The contour interval for
  $A+B_0 y$ is $0.005$, $0.02$ and $0.08$ (left to right).}
\label{fig:q512g}
\end{figure}

\begin{figure}
\begin{center}
\includegraphics[width = \textwidth]{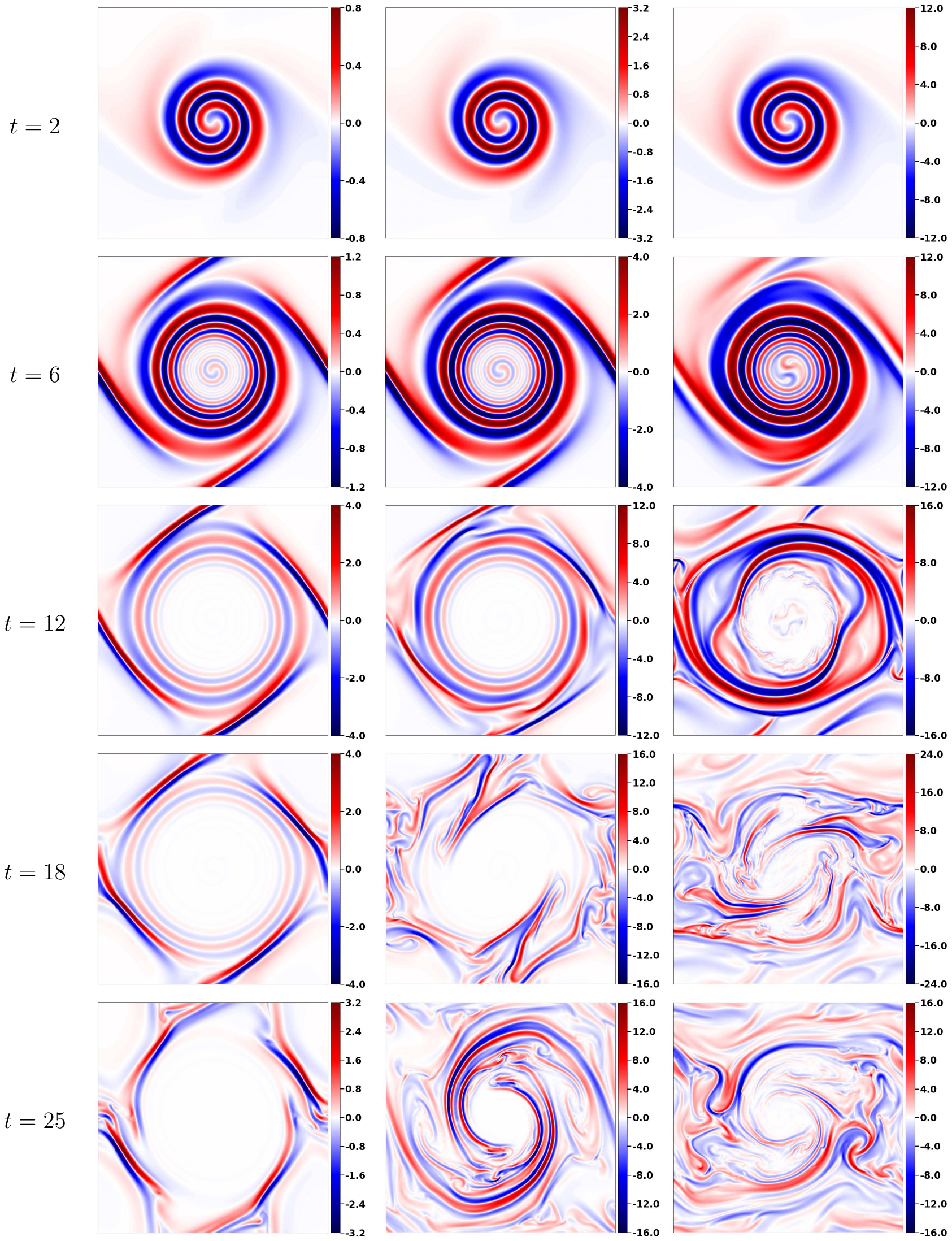}
\end{center}
\caption{As in figure~\ref{fig:q512g} but for the current density
  $j(\bx,t)$.}
\label{fig:j512g}
\end{figure}

We begin by discussing a few characteristic simulations.  For this
purpose, we consider $\delta=0.025$ (corresponding to $n_g=512$) and
three different values of $\gamma$, namely $0.125$, $0.5$ and $2$,
corresponding to a weak, moderate and strong magnetic field (at full
intensity).  For the initially Gaussian vortex, figure~\ref{fig:q512g}
shows the evolution of the vorticity field (colour) overlaid with
field lines (contours of $A+B_0 y$) for the three values of $\gamma$
(increasing from left to right).  The corresponding current density
is shown in figure~\ref{fig:j512g}.

In the left column, the weak field has no discernible impact on the
vortex, which remains approximately circular and undiminished
throughout the evolution.  The field itself is twisted rapidly and
field lines are broken by the magnetic diffusivity, leading to an
expanding region of nearly zero field --- this is the classic `flux
expulsion' phenomenon first described by \cite{weiss_1966}.  Note that
the current sheets and strong field gradients developing in the
periphery are a consequence of periodicity; otherwise the region of
nearly zero field would continue expanding.  These strong gradients
generate vorticity which subsequently destabilises and disrupts
the field in this region, away from the central vortex.

In the middle column for a stronger initial field, the initial
evolution is closely similar.  Now however a weak spiral pattern in
vorticity is generated by the tightening field gradients before
diffusion erases them.  Moreover, the outer gradients induced by
periodicity are much stronger and more disruptive at late times,
distorting the vortex into an elliptical shape and eroding it.  The
expelled field in this case collapses back toward the vortex and
strongly interacts with it at late times --- again here the later dynamics
is a consequence of periodicity.

In the right column for the strongest initial field, again the early
evolution is similar.  Also, spiral vorticity generation occurs but
this time the vorticity becomes comparable with that in the core, and
the alternating bands of vorticity destabilise by $t=12$.  The
subsequent evolution is much more turbulent and noticeably affects the
vortex core itself, causing strong distortion and a loss of symmetry
at late times.  In this case the field is never fully expelled
initially and the strong gradients collapse back more rapidly, leaving
weaker gradients only in the vortex core.  A slightly higher initial
field strength ($\gamma=2.5$) leads to the complete destruction of the vortex
by $t=25$ (not shown).

\begin{figure}
\begin{center}
\includegraphics[width = \textwidth]{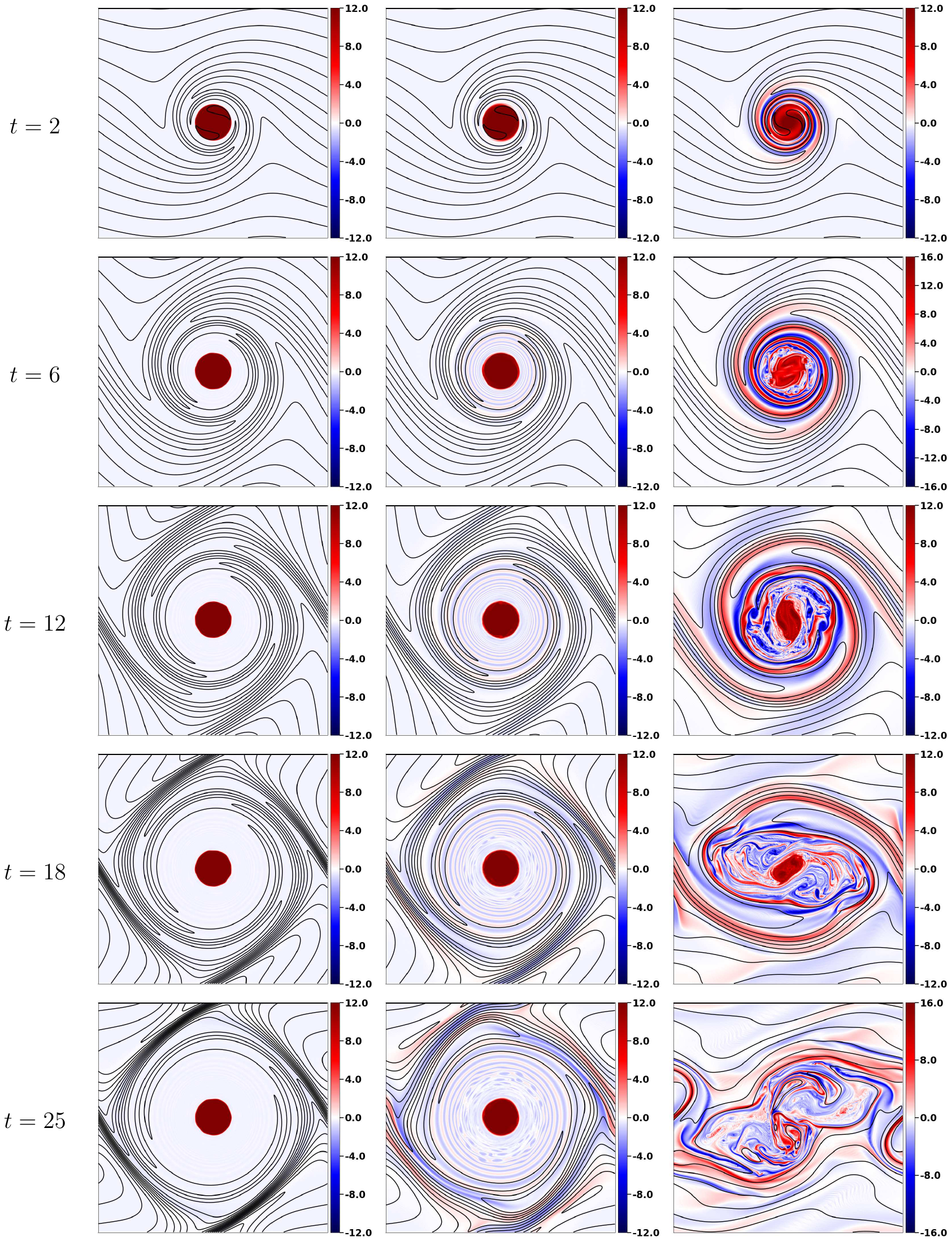}
\end{center}
\caption{Vorticity $\omega(\bx,t)$ and magnetic field lines at the
  times indicated (top to bottom) for an initially uniform (Rankine) vortex and
  for $\gamma=0.125,\,0.5$ and $2$ (left to right).  The vorticity
  colourbar is indicated next to each image.  The contour interval for
  $A+B_0 y$ is $0.005$, $0.02$ and $0.08$ (left to right).}
\label{fig:q512p}
\end{figure}

\begin{figure}
\begin{center}
\includegraphics[width = \textwidth]{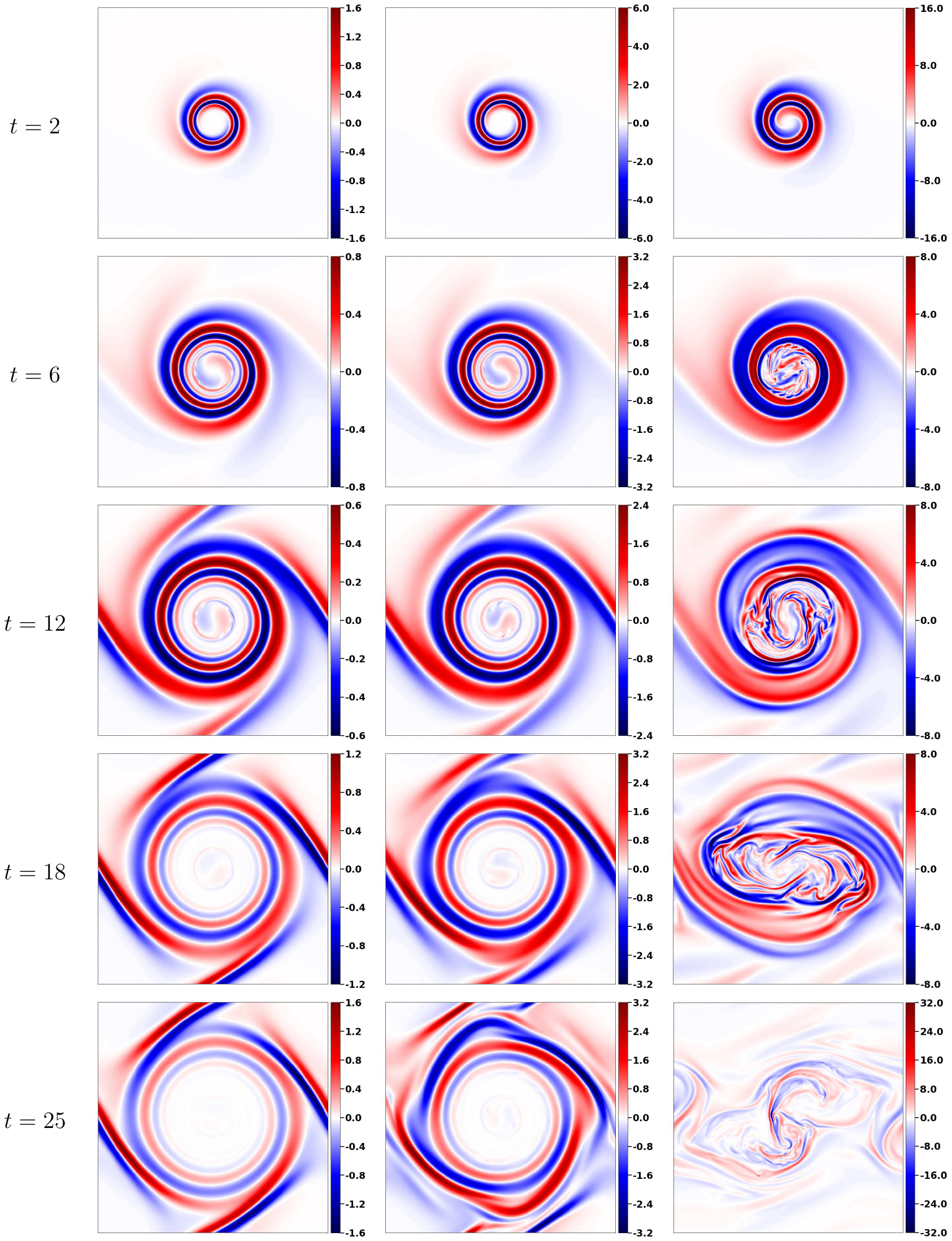}
\end{center}
\caption{As in figure~\ref{fig:q512p} but for the current density
  $j(\bx,t)$.}
\label{fig:j512p}
\end{figure}

We now contrast the Gaussian vortex with a circular vortex patch of
uniform initial vorticity.  We choose the same $\delta$ and $\gamma$
values so that any differences may be attributed to the initial
vorticity profile.  The vorticity (with superimposed field lines) and
current density are illustrated in figures~\ref{fig:q512p} and
\ref{fig:j512p} respectively.  Compared with figures~\ref{fig:q512g} and
\ref{fig:j512g}, there are striking differences.  First of all, in
general, the magnetic field amplification is much stronger for the patch
than for the Gaussian vortex, leading to greater disruption at late
times and indeed vortex destruction for $\gamma=2$.  In that case, the
field is never expelled and interacts strongly with the vortex,
ultimately pulling it apart and leaving intense current sheets where
the vortex used to be.  For moderate initial field strength
($\gamma=0.5$, middle column), the vorticity filament spiral outside
of the initial vortex destabilises, something only seen for $\gamma=2$
in the case of the Gaussian vortex.  Even for weak initial field ($\gamma=0.125$, left column), there is evidence that the
field is disturbing the vortex boundary, which is no longer circular
but more polygonal.  Evidently, the less regular flow associated with
the vortex patch, especially the discontinuity in shear at its edge,
gives rise to a much stronger local interaction with the magnetic
field.  This is quantified below.

\subsection{Quantitative results}

We next discuss various quantitative aspects of the flow evolution
before developing a scaling theory to explain the results.  An
important diagnostic is the vortex circulation $\Gamma$ defined in
(\ref{circ}), in particular its rate of change $d\Gamma/dt$, which may
be computed using (\ref{dcirc}).  The circulation changes only as a
result of the magnetic field, and given that field lines are twisted
by the vortex, we expect {\it a priori} that the associated tension
will act to slow the vortex rotation, reducing its circulation.  At
least this should happen initially.

\begin{figure}
\begin{center}
\includegraphics[width = \textwidth]{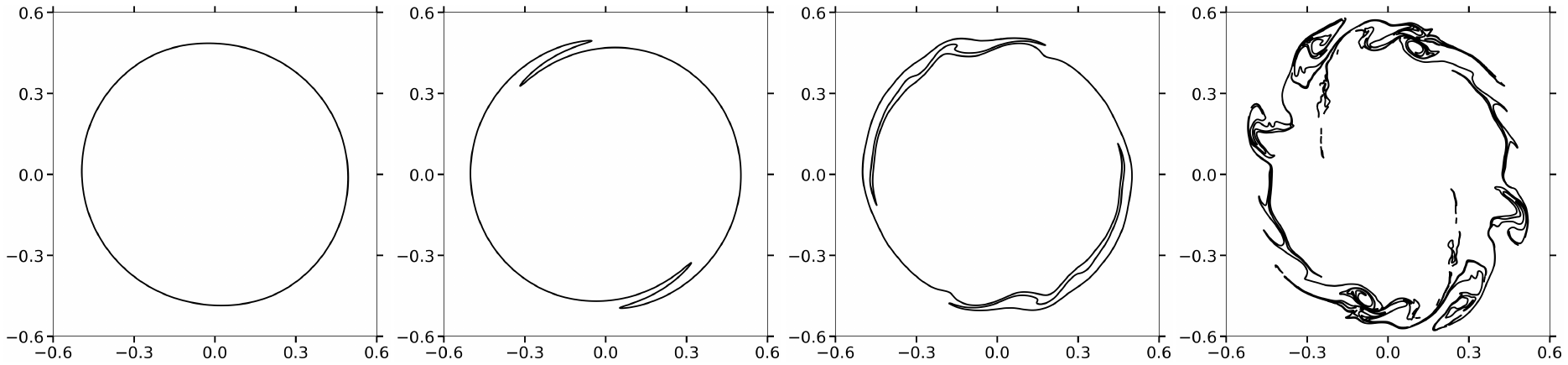}
\end{center}
\caption{The material contour (or contours) belonging to the
  original circular tracer contour at times $t=5,\,10,\,15$ and
  $20$ for the Gaussian vortex case with $\delta=0.0125$
  ($n_g=1024$) and $\gamma=2$.  Only the inner portion of the
  flow domain is shown.  }
\label{fig:tracer}
\end{figure}

We compute the circulation using a material contour ${\mathcal{C}}$
which is initially a circle of radius $R$ centred at the origin.  The
contour is evolved in just the same way as the vorticity contours,
except that it is never rebuilt.  It may generally distort and even
split into various parts, but at all stages the collection of contours
belonging to the original contour is used in the contour integrations
required in (\ref{circ}) and (\ref{dcirc}).  A more extreme example
showing how ${\mathcal{C}}$ may distort is provided in
figure~\ref{fig:tracer}.  Here, at late times, ${\mathcal{C}}$ splits
into many parts, though most of these are very small.  Nonetheless,
accurate contour integration can be performed around such contours.
Here, we use two-point Gaussian quadrature and the actual curved shape
of the contour between successive nodes.  At the evaluation points,
the fields such as $\bu$ and $j$ are interpolated from nearby gridded
values using bi-linear interpolation.

\begin{figure}
\begin{center}
\includegraphics[width = \textwidth]{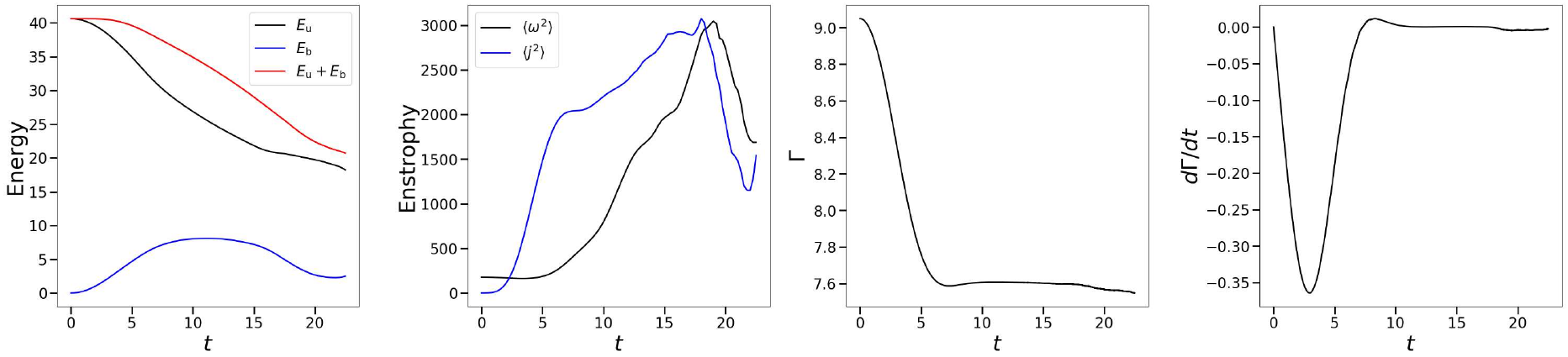}
\end{center}
\caption{Time evolution of various diagnostics for the Gaussian vortex
  case with $\delta=0.0125$ and $\gamma=2$, i.e.\ as in
  figure~\ref{fig:tracer}.  From left to right, we show energy,
  enstrophy, circulation and the circulation rate of change.  See text
  for details.}
\label{fig:diag}
\end{figure}

Figure~\ref{fig:diag} shows $\Gamma$ and $d\Gamma/dt$ along with other
important diagnostics for an initially Gaussian vortex when
$\delta=0.0125$ and $\gamma=2$.  This figure also shows the
hydrodynamic (or kinetic) and magnetic energy components,
respectively
\beq
\label{energy}
E_u=\frac{1}{2}\int_{-\pi}^{\pi}\int_{-\pi}^{\pi}|\bu|^2 dx\,dy
\qquad\textrm{and}\qquad
E_b=\frac{1}{2}\int_{-\pi}^{\pi}\int_{-\pi}^{\pi}|\nabla{A}|^2 dx\,dy
\eeq
together with the total energy $E=E_u+E_b$, as well as the
hydrodynamic and magnetic `enstrophies' $\langle\omega^2\rangle$ and
$\langle j^2\rangle$, here defined as the domain mean values of
$\omega^2$ and $j^2$.  Note that $E_b$ does not include the constant
part of the magnetic energy associated with the initial mean magnetic
field $B_0\bex$.  For the hydrodynamic case the enstrophy $\langle\omega^2\rangle$ is conserved in the absence of viscosity, whilst for the ideal MHD case the total energy $E$ (magnetic plus kinetic) is conserved, whilst the enstrophy is not.
Note, in this limit $\langle j^2\rangle$ is
not conserved, however any functional of $A+B_0 y$ is, since this field
is materially conserved (and the flow is incompressible).  However,
this limit is not relevant for the present purposes and is likely to
be mathematically ill-posed \citep{cao_2017}.

The case illustrated in figure~\ref{fig:diag} is broadly
representative of all simulations conducted.  All share similar trends
albeit with different amplitudes and time scales.  The kinetic energy
decays while the magnetic energy grows non-diffusively at first, as
evidenced by the conservation of the total energy at early times (red
curve).  As field gradients increase, magnetic diffusion
and the Lorentz force begin
 to act, ultimately halting the growth in magnetic energy and inducing
decay at later times.  Note that the background magnetic energy
$\bar{E}_b=\tfrac12 B_0^2 (2\pi)^2$, which is ignored in $E_b$, is
typically small compared with the maximum $E_b$ observed.  In this
example, $\bar{E}_b\approx 0.117$, which is less than $2\%$ of the
maximum $E_b$.  The decay in kinetic energy $E_u$ is consistent with
the action of the field, forcing the vortex to slow down as 
field lines twist.  Only the breaking of field lines by the diffusion
allows the vortex to survive.

Regarding the enstrophy and current density, the mean square current density
$\langle j^2\rangle$
initially grows rapidly until approximately $t=6$, during which time
the hydrodynamic enstrophy $\langle\omega^2\rangle$ weakly decays.
Over this period, the vortex is mainly slowing down as it twists
magnetic field lines.  Around $t=6$, magnetic diffusion begins to act
strongly, expelling the field from the vortex core and its immediate
vicinity.  After this time, $\langle j^2\rangle$ grows more slowly but
$\langle\omega^2\rangle$ begins to grow rapidly, reaching a value more
than $17$ times its initial value by $t=19$ before subsequently
decaying.  This growth is due to the vorticity production by the
strong current sheets created in the periphery of the vortex,
primarily around and after $t=6$ (cf.\ right column in
figure~\ref{fig:j512g}).  Only when these sheets begin to decay
significantly at late times does $\langle\omega^2\rangle$ begin to
decay.  This decay is not inviscid: numerical dissipation acting at
scales well below the grid scale are sufficient to cause a strong
decay when $\omega$ develops extensive fine-scale structure in the
form of filaments or sheets.

The time evolution of the circulation $\Gamma$ and its rate of change
$d\Gamma/dt$ are consistent with the foregoing description.  Up to
$t=6$, the circulation continually decreases as the twisting magnetic
field lines slow the vortex rotation.  The slowing decay after $t=3$
indicates that magnetic diffusion is already breaking field lines and
thus reducing the impact on the vortex.  After $t=6$, the circulation
remains roughly constant (with a small rebound just after $t=6$),
despite the continued vorticity production beyond the vortex core.
This is because the contour ${\mathcal{C}}$ around which the
circulation is computed remains near the centre of the domain, as
shown previously in figure~\ref{fig:tracer}.  Notably, since the
circulation of the entire domain must be zero in a doubly-periodic
domain, then the total circulation outside of ${\mathcal{C}}$ is just
$-\Gamma$.  In particular, this means that despite the intense
vorticity production occurring beyond the vortex core, the net change
in total vorticity there is small after $t=6$.

\subsection{Scaling theory}

\begin{figure}
\begin{center}
\includegraphics[width = \textwidth]{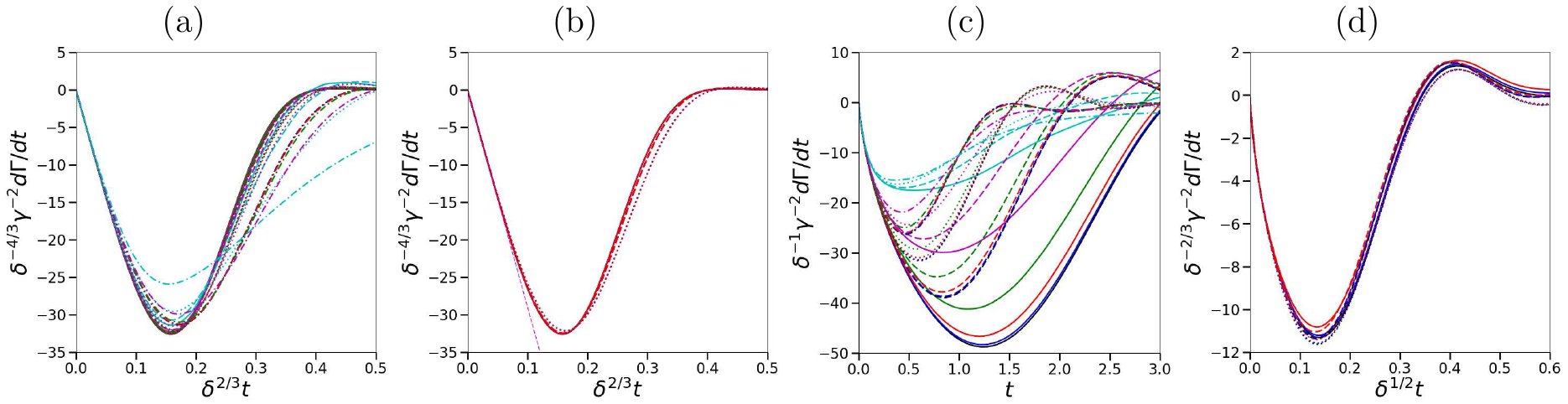}
\end{center}
\caption{Scaled forms of $d\Gamma/dt$, as indicated, for
  $\delta=0.0125,\,0.025,\,0.05,\,0.1$ and
  $\gamma=0.0625,\,0.125,\,0.25,\,0.5,\,1,\,2$.  The values of
  $\delta$ are distinguished by line style: solid, dashed, dotted,
  dash-dotted.  The values of $\gamma$ are distinguished by colour:
  black, blue, red, green, magenta and cyan.  Panels (a) and (b)
  pertain to the initially Gaussian vortex, while panels (c) and (d)
  pertain to the initially circular vortex patch.  The straight dashed
  magenta line in panel (b) is the fit to equation
  (\ref{gauss-dcirc-bis}).  See text for further details.}
\label{fig:scaling}
\end{figure}

We next focus on the behaviour of $d\Gamma/dt$, in particular how its
evolution depends on the parameters $\delta$ and $\gamma$ as well as
on the initial vorticity distribution.  Given the crucial role played
by the magnetic diffusivity, we expect the flow to evolve on an
appropriate diffusive time scale, which may (and does) depend on the
initial vorticity distribution.  Moreover, the amplitude of
$d\Gamma/dt$ is expected to increase as $\delta$ and $\gamma$
increase.  Figure~\ref{fig:scaling} summarises all of the results,
both for the Gaussian vortex (panels (a) \& (b)) and for the circular
vortex patch (panels (c) \& (d)), for 4 values of
$\delta\in[0.0125,0.1]$ differing by factors of two, and for 6 values
of $\gamma\in[0.0625,2]$ also differing by factors of two.

First consider panels (a) and (b) for the initially Gaussian vortex.
Panel (b) merely shows a restricted set of parameters, namely the
three smallest $\delta$ values and the three smallest $\gamma$ values.
This is done to show just how well the curves collapse for small
$\delta$ and $\gamma$, where one might expect a scaling theory to
apply.  The results show a clear dependence on the scaled time
$\delta^{2/3}t$, rather than the $\delta^2 t$ dependence one might
expect for pure diffusive decay.  This is due to the spiral wind-up of
the magnetic field (which at small $\gamma$ behaves like a passive
tracer), accelerating the diffusive decay ($\propto\eta^{1/3}$), as
originally shown by \cite{bbg_2001}.  The quadratic dependence of
$d\Gamma/dt$ on $\gamma$ is just a consequence of the form of the
integrand in (\ref{dcirc}), where a quadratic product of quantities
proportional to the magnetic field appears.

The dependence on $\delta^{4/3}$ may be explained by considering the
early time evolution of $d\Gamma/dt$ and using Moffatt's approximate
inviscid solution for $A$,
\beq
\label{moffatt}
A(r,\theta,t)=B_0 r [\sin(\theta-\bar\Omega t) - \sin\theta]
\eeq
\citep{moffatt_1983}. 
This assumes that the background flow is steady and axisymmetric, with
angular frequency $\bar\Omega(r)$ a function of $r$ only, and moreover
that $\gamma\ll 1$.  This solution directly follows from
(\ref{poteq}), neglecting $\eta$, and can be seen most easily by
recognising that the total potential $\tilde{A}=A+B_0 y$ satisfies
(for a steady axisymmetric flow)
\beq
\tilde{A}_t+\bar\Omega\tilde{A}_\theta=0,
\eeq
implying $\tilde{A}=F(r,\theta-\bar\Omega t)$ for some function $F$.
This function is determined from the initial conditions,
$\tilde{A}(r,\theta,0)=B_0 y = B_0 r\sin\theta$, giving
$F(r,\theta)=B_0 r\sin\theta$, from which (\ref{moffatt}) follows.
Now, to estimate $d\Gamma/dt$ at early times (before diffusion has
a chance to act significantly), we also need the current density $j$.
Using Moffatt's solution (\ref{moffatt}) for $A$, we find
\beq
\label{jmoffatt}
j(r,\theta,t)=B_0 t(r\bar\Omega_{rr}+3\bar\Omega_r)\cos(\theta-\bar\Omega t)
             +B_0 t^2 r\bar\Omega_r^2\sin(\theta-\bar\Omega t).
\eeq
Using this and $\tilde{A}_\theta=B_0 r\cos(\theta-\bar\Omega t)$ in
(\ref{dcirc}) written using the parametrisation $\theta$, we obtain
after elementary integration
\beq
\label{gauss-dcirc}
\frac{d\Gamma}{dt}=B_0^2 \pi t (r^2\bar\Omega_{rr}+3r\bar\Omega_r)_{r=R}
\eeq
(the $t^2$ term integrates to zero).  But for the Gaussian vortex,
\beq
\label{gauss-omega}
\bar\Omega(r)=\omega_{\mathsf{max}}\frac{R^2}{r^2}
\left(1-e^{-r^2/2R^2}\right).
\eeq
After some algebra, one may show that
\beq
\label{nasty-term}
(r^2\bar\Omega_{rr}+3r\bar\Omega_r)_{r=R}=-\omega_{\mathsf{max}}e^{-1/2}.
\eeq
Next, using $B_0=\delta\gamma U_0=\tfrac12\delta\gamma\omega_0 R$ 
from (\ref{nondim}) in (\ref{gauss-dcirc}) together with 
$\omega_{\mathsf{max}}=\omega_0/(2(1-e^{-1/2}))$, we obtain the
following explicit prediction for $d\Gamma/dt$:
\beq
\label{gauss-dcirc-bis}
\frac{d\Gamma}{dt}=-\delta^2\gamma^2\frac{\pi\omega_0^3 R^2}
                                         {8\left(e^{1/2}-1\right)} t.
\eeq
This is plotted as the straight dashed magenta line in panel (b).  It
coincides with the initially linear decrease in $d\Gamma/dt$.
Moreover, it shows that $\delta^{-4/3}\gamma^{-2}d\Gamma/dt$ should be
proportional to the scaled time $\delta^{2/3}t$, as observed.  At
later times, $d\Gamma/dt$ is affected by magnetic diffusion, neglected
here.  This arrests the change in circulation, ultimately reducing it
to near zero at late times.
This scaling is consistent with the early-time behaviour
described by \citet{gmt_2016}.

Next consider panels (c) and (d) for the circular vortex patch.  In
(c), all of the results are plotted versus the unscaled time $t$ but
$d\Gamma/dt$ is scaled in such a way that the very early time
behaviour is closely similar: all 24 curves lie on top of each other
for $t<0.2$.  In panel (d), we show that by plotting $d\Gamma/dt$
versus the scaled time $\delta^{1/2}t$ and appropriately scaling
$d\Gamma/dt$, the curves nearly collapse onto a single curve.  Here
again only the three smallest $\delta$ values and the three smallest
$\gamma$ values are plotted.  The collapse is not as good as for the
Gaussian vortex in panel (b), but nonetheless extends well into the
evolution, including the rebound after $\delta^{1/2}t=0.35$ (which is
much stronger than seen for the Gaussian vortex).

From the earliest time, diffusion plays an important role, and hence the
inviscid solution presented above for the Gaussian vortex does not
apply.  Equation (\ref{gauss-dcirc}) is problematic for the vortex
patch, as the radial function there equals $-\omega_0 R \delta(r-R)$,
where here $\delta(s)$ is Dirac's delta function.  In reality, $j$
immediately diffuses, and qualitatively should exhibit a
$1/\sqrt{\eta{t}}$ time dependence at $r=R$ based on the solution to
the heat equation.  This would imply
\beq
\label{patch-dcirc}
\frac{d\Gamma}{dt} \sim -B_0^2 t \frac{\omega_0 R}{\sqrt{\eta t}}
                   \sim -\delta\gamma^2 \omega_0^2 R^2 \sqrt{\omega_0 t}
\eeq
(the constant of proportionality based on the solution of the heat
equation is $\sqrt{\pi}/8$).  While this qualitatively explains the
observed nonlinear time dependence in $d\Gamma/dt$ in panels (a) and
(b), it represents only a crude fit.  This is likely due to assuming
that $j$ evolves according to the heat equation while using the
inviscid solution for $A$.  Nonetheless, (\ref{patch-dcirc}) explains
the dependence on $\delta\gamma^2$ exhibited in panel (c).

\begin{figure}
\begin{center}
\includegraphics[height = 0.333\textwidth]{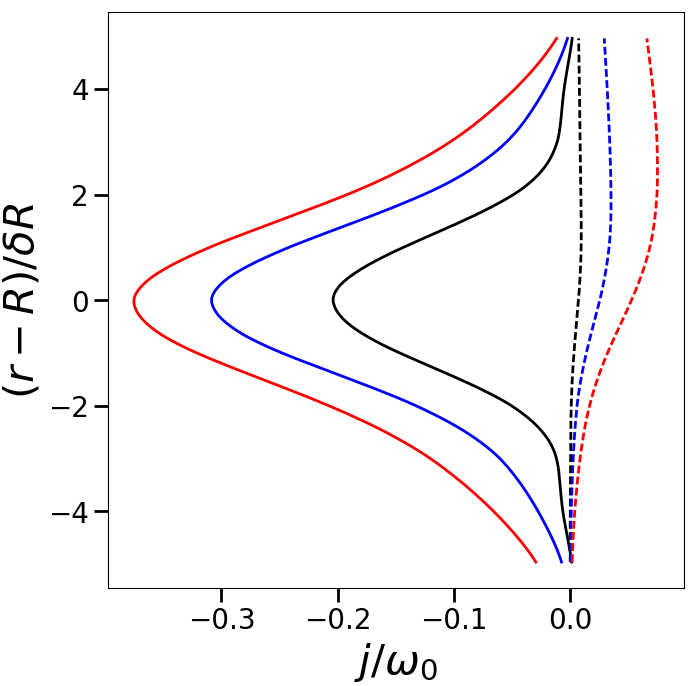} 
\includegraphics[height = 0.333\textwidth]{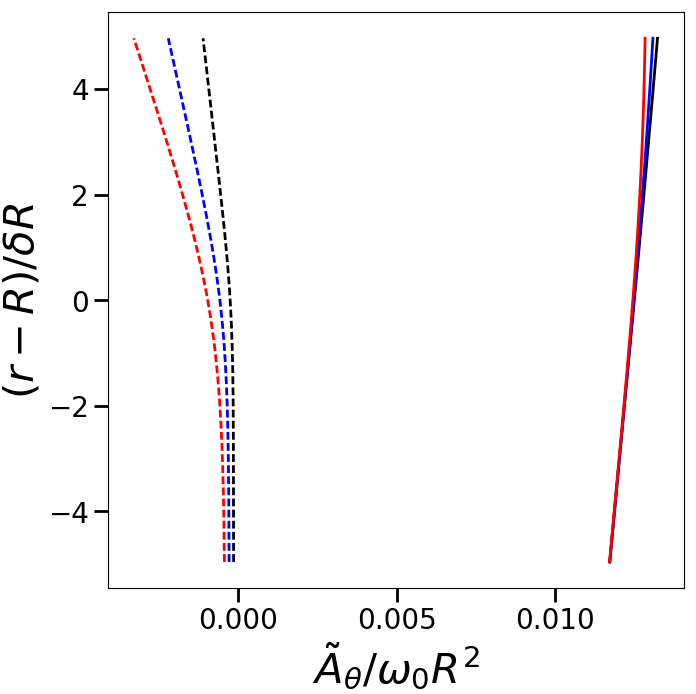}
\end{center}
\caption{Scaled radial dependencies of $j$ and $\tilde{A}_\theta$
  near the edge of a circular vortex patch, projected onto the
  cosine and sine azimuthal modes (solid and dashed respectively)
  at times $t=0.1$ (black), $0.2$ (blue) and $0.3$ (red).  These
  results are derived from a simulation starting from a circular
  vortex patch with $\delta=0.0125$ and $\gamma=1$.}
\label{fig:cross}
\end{figure}

The actual radial cross sections of $j$ and $\tilde{A}_\theta$ near
the vortex edge, projected onto the $\cos(\theta-\bar\Omega t)$ and
$\sin(\theta-\bar\Omega t)$ azimuthal modes, are shown in
figure~\ref{fig:cross}.  We find that the cosine projection of
$j$ indeed exhibits a roughly Gaussian dependence on $r-R$, while
the cosine projection of $\tilde{A}_\theta$ varies only weakly
across $r=R$.  However, the sine projection of $\tilde{A}_\theta$
is a diffusive effect, and spoils any simple theoretical prediction
for $d\Gamma/dt$.

Finally, we remark that $|d\Gamma/dt|$ grows to much larger values for
the initially circular patch than for the Gaussian.  As an example,
for the smallest $\delta$ and $\gamma$ values, the ratio in the
maximum values of $|d\Gamma/dt|$ is $6.45$.  The circular patch
therefore is much more strongly affected by the magnetic field
than the Gaussian vortex.

\begin{figure}
\begin{center}
\includegraphics[height = 0.333\textwidth]{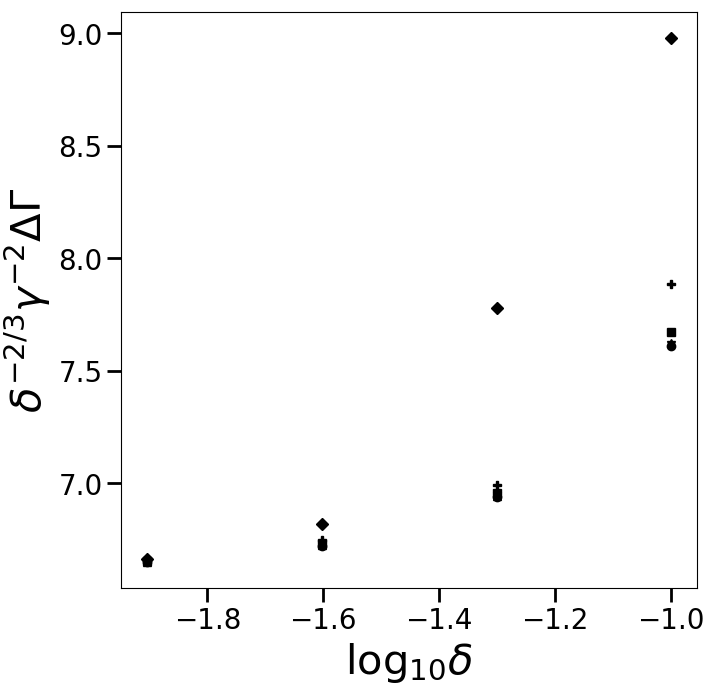} 
\includegraphics[height = 0.333\textwidth]{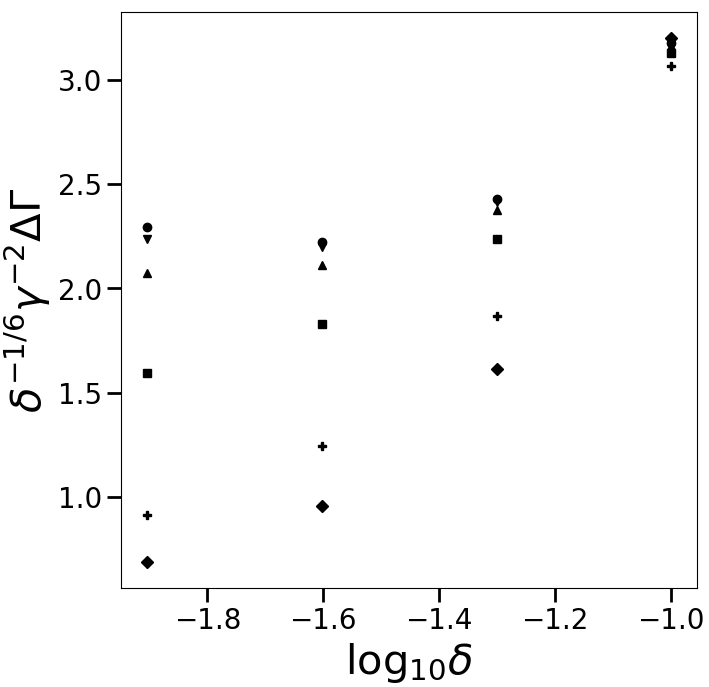}
\end{center}
\caption{Scaled net reduction in circulation (a) for the initial
  Gaussian vortex and (b) for the initial circular vortex patch,
  as a function of $\delta$.  The symbols indicate the value of
  $\gamma$.  In order of increasing $\gamma$, the symbols are
  circle, triangle pointing down, triangle pointing up, square,
  plus and diamond.}
\label{fig:netcirc}
\end{figure}

In nearly all simulations conducted for either vortex profile, the
circulation $\Gamma$ decreases and levels off at a nearly constant
value, particularly for small $\delta$ and $\gamma$ (when the vortex
is not greatly disrupted).  The above scaling results can be used to
estimate the total reduction in circulation $\Delta\Gamma$.  For the
initially Gaussian vortex, integrating $-d\Gamma/dt$ from $t=0$ to
$t\sim\delta^{-2/3}$ gives the estimate
$\Delta\Gamma\sim\delta^{2/3}\gamma^2$.  This is shown to predict well
the actual change in circulation up to $t=0.5\delta^{-2/3}$, as shown
in panel (a) of figure~\ref{fig:netcirc}.  We stress that, in order to
assess the ultimate role of the magnetic field in extracting
circulation from the vortex --- and therefore the ultimate fate of the
vortex, the {\it integrated} effects of the
Lorentz force must be calculated. The same integration for the patch
to $t=0.6\delta^{-1/2}$ gives the estimate
$\Delta\Gamma\sim\delta^{1/6}\gamma^2$.  As seen in panel (b), this is
not accurate, and suggests that the dependence on $\gamma$ should be
different.  However, by log-scaling the data and searching for best
fit curves, no simple relationship was found to collapse the data
significantly better than shown in panel (b).  The circular vortex
patch is not nearly as simple as the Gaussian vortex.


\subsection{Initially elliptical vortex patches}

\begin{figure}
\begin{center}
\includegraphics[width = \textwidth]{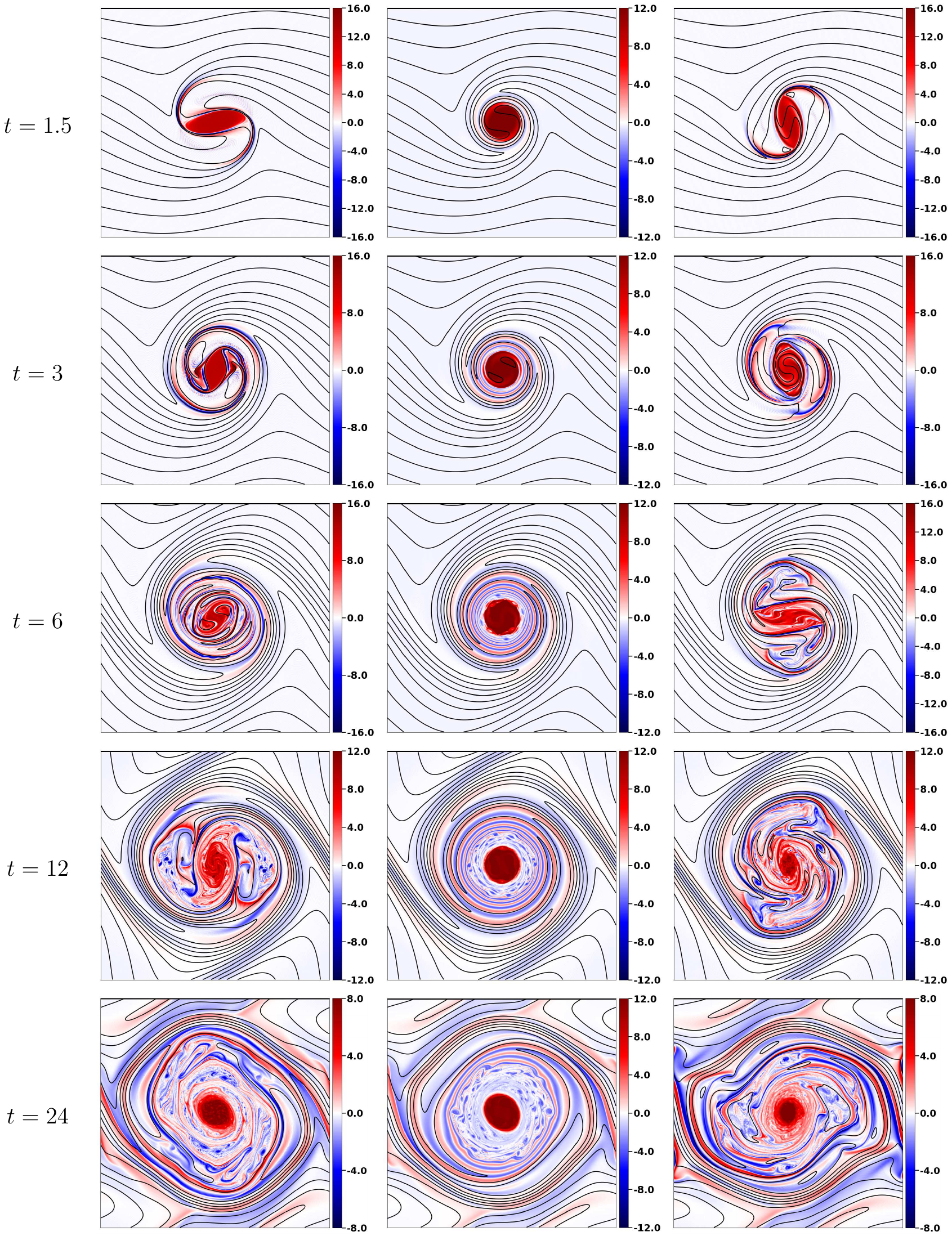}
\end{center}
\caption{Vorticity $\omega(\bx,t)$ and magnetic field lines at the
  times indicated (top to bottom) for an initially elliptical vortex
  with $\lambda=1/3,\,1$ and $3$ (left to right).  The vorticity
  colourbar is indicated next to each image.  The contour interval for
  $A+B_0 y$ is $0.04$ in all images.}
\label{fig:q512e}
\end{figure}

\begin{figure}
\begin{center}
\includegraphics[width = \textwidth]{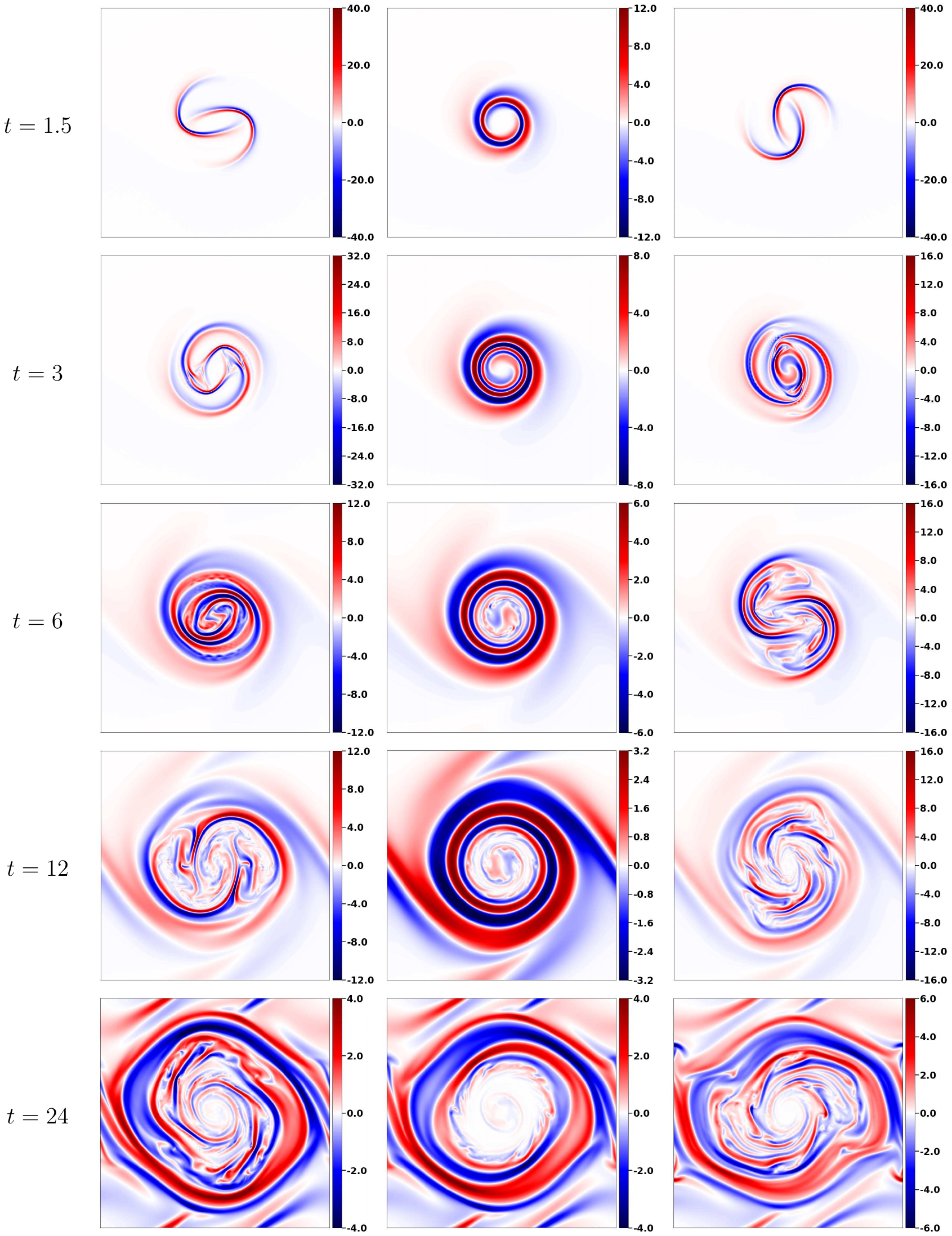}
\end{center}
\caption{As in figure~\ref{fig:q512e} but for the current density
  $j(\bx,t)$.}
\label{fig:j512e}
\end{figure}

We conclude this section by discussing the case of an initially
elliptical vortex patch whose boundary is given by
\beq
\label{epatch}
\frac{x^2}{a^2}+\frac{y^2}{b^2}=1
\eeq
with $\sqrt{ab}=R$ so that it has the same area for all aspect ratios
$\lambda=b/a$.  One reason to study the ellipse is to understand the
role of the threading of the field lines in the subsequent evolution
of the vortex \citep[see e.g.][]{KSD_2008}.
Would a vortex threaded by more field lines
($\lambda>1$) be more affected by the magnetic field than one which is
threaded by fewer field lines ($\lambda<1$)?  As a first qualitative
view, figures~\ref{fig:q512e} and \ref{fig:j512e} show the vorticity,
field lines and current density for three ellipses, with $\lambda=1/3$
(left), $\lambda=1$ (middle), $\lambda=3$ (right) --- all for
$\delta=0.025$ and $\gamma=1$.  Note that an ellipse with
$\lambda<1/3$ or $\lambda>3$ is linearly unstable in the absence of a
magnetic field \citep{love_1893}.  Here the field acts to circularise
the vortex.  For both $\lambda=1/3$ and $\lambda=3$, the impact of the
field on the vortex is noticeably greater than for the circular
vortex, $\lambda=1$.  It appears that $\lambda=3$ is most strongly
affected, but the argument that more field lines initially
threading the vortex
has a greater impact is clearly not correct.  The circular vortex is
much less affected than the vortex with $\lambda=1/3$, which is
threaded by the fewest field lines.

\begin{figure}
\begin{center}
\includegraphics[width = \textwidth]{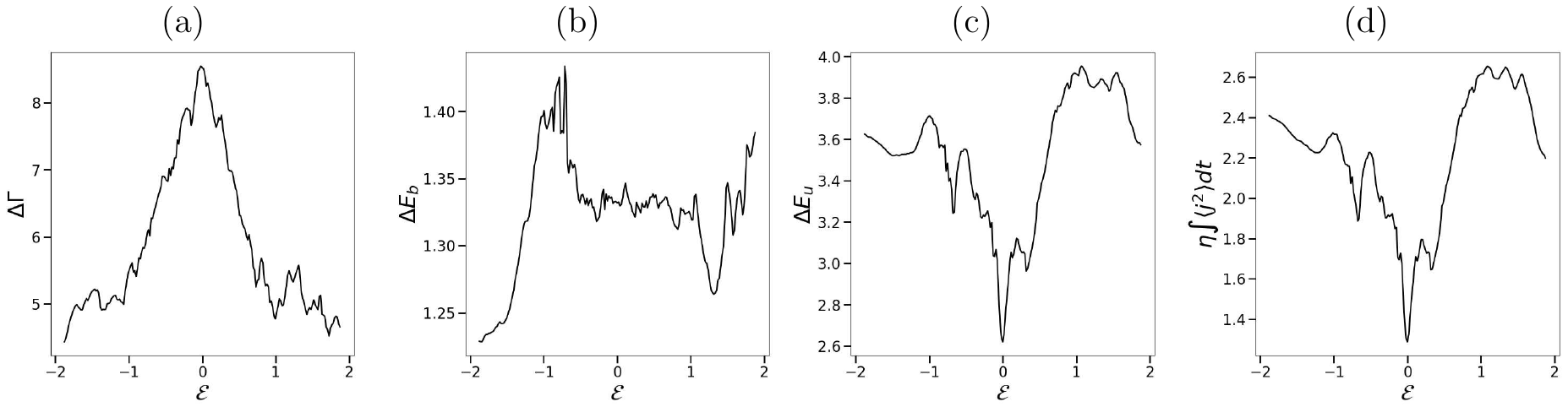}
\end{center}
\caption{Accumulative diagnostics as a function of initial
  vortex eccentricity ${\mathcal{E}}$.  The diagnostics are
  computed over $0\leq t \leq 10$ and show (a) the net circulation
  reduction $\Delta\Gamma$, (b) the net growth in magnetic energy
  $\Delta{E_b}$, (c) the net decline in kinetic energy 
  $\Delta{E_u}$, and (d) the net magnetic dissipation
  $\eta\int_0^{10}\langle{j^2}\rangle\,dt$.  All simulations
  used the parameters $\delta=0.025$ and $\gamma=1$.}
\label{fig:ellipse}
\end{figure}

Instead, what appears to be important is the reach of the vortex in
sweeping through the background magnetic field.  Then, both small and
large $\lambda$ would exhibit a similar behaviour, and would be more
affected than the circular vortex.  This is borne out in
figure~\ref{fig:ellipse}, which plots various accumulative diagnostics
for over 200 simulations varying the vortex eccentricity
\beq
\label{eccen}
{\mathcal{E}}=\frac12\left(\lambda-\frac{1}{\lambda}\right)
\eeq
over the range $-15/8 \leq {\mathcal{E}} \leq 15/8$, corresponding to
$1/4 < \lambda < 4$.  The results in figure~\ref{fig:ellipse} show that
the circular vortex is clearly anomalous.  Yet, the circular vortex
also displays the greatest net reduction in circulation, despite
keeping it shape.  Ellipses retain their initial circulation better,
not because they are less affected, but because the vorticity
production near the vortex edge is different: both increases and
decreases in vorticity occur within the vortex edge, whereas for a
circular vortex only decreases occur.  This can be seen to some extent
in figure~\ref{fig:q512e}, where the plotted vorticity field levels
are higher for both $\lambda=1/3$ and $\lambda=3$ than for
$\lambda=1$.  In short, the circulation changes occurring for
distorted vortices are more complex but tend to be weaker than those
occurring for an initially circular vortex.

Regarding the other diagnostics, there is very little dependence in
the gain in magnetic energy on vortex eccentricity, but a significant
dependence in the loss in kinetic energy.  The initially circular
vortex exhibits the least loss in kinetic energy, consistent with the
images in figure~\ref{fig:q512e}, where the flow surrounding the
vortex is generally much less agitated compared with the elliptical
cases.  Finally, the net magnetic dissipation is also weakest for the
initially circular vortex, consistent with the generally weaker field
of current density seen in the middle panels of
figure~\ref{fig:j512e}.

Finally, despite the fact that the ellipse is unstable for
$\lambda<1/3$ or $\lambda>3$ in the absence of a magnetic field, the
effect of the magnetic field in the examples shown (which here is not a
weak effect) dominates the flow evolution, rapidly causing the vortex to
become more circular in form.  A weaker magnetic field might allow a
competition between the hydrodynamic instability and the action of the
field, but this is beyond the scope of the present study.

\section{Conclusions and Future Work}

In this paper we have examined the role of a weak magnetic field in
modifying Kelvin's circulation theorem in two-dimensional MHD at low
magnetic Prandtl number $Pm$. The circulation of the velocity field is
a materially conserved quantity in hydrodynamics and plays a key role
in determining the dynamics. That the magnetic field acts through the
Lorentz force to destroy these conservation properties is well known,
but quantifying the effect as a function of parameters (magnetic
diffusion and field strength) has hitherto not been achieved. We
consider three model flows that, in the absence of magnetic effects,
remain exact stable solutions of the two-dimensional Euler equations;
namely the Gaussian vortex, the circular vortex patch, and the
elliptical vortex patch.

As always in these situations, it appears as though the role of the
magnetic field is more subtle than first (or even second) imagined. As
noted by \citet{gmt_2016} the degree of circulation extracted from the
vortices {\it must} be determined via integrating the effects of the
Lorentz force over the entirety of the evolution, rather than by
hypothesising balances of crude instantaneous measures of the Lorentz
force, which will yield incorrect scalings. Furthermore, the roles of
diffusion and Lorentz force depend on the initial configuration of the
flow, with smooth vortices reacting rather differently from vortex
patches. 

We note that the dynamics and role of the magnetic field is different at high $Rm$ fluids at low $Pm$ (such as astrophysical fluids) than it is for low $Rm$ fluids at low $Pm$ such as liquid metal experiments, and even some geophysical cases. The latter situation
can be investgated in detail, as then the Lorentz force that breaks Kelvin's theorem can be linearised about a background magnetic field. We are currently investigating the scalings that apply.

Although we have focussed on simple models with no background rotation
or stratification, it is clear that the results can (and should) be
extended to more geophysically and astrophysically relevant cases with
these included. Simple extensions that should be (and are being)
evaluated include incorporating rotation and stratification on
both a $\beta$-plane and a spherical surface. Here the key materially
conserved quantity is the absolute (potential) vorticity, which
includes the planetary and relative vorticity as well as density
variations. Since material conservation of this scalar field has a
significant effect on the dynamics (for example being implicated in
the formation of potential vorticity staircases, see
e.g.\ \cite{DM:2008}), then understanding the role of a magnetic field
in modifying this conservation property is an important next step in
understanding the dynamics of stably stratified magnetised
environments \citep[as described in][]{T_2005}.



\section*{Acknowledgements}
PHD and SMT would like to acknowledge the support of the Festival de Th\'eorie in Aix-en-Provence. Support for this research has come from the UK Engineering and Physical Research Council (grant nu. EP/H001794/1).


%
\bibliographystyle{jfm}
\bibliography{2DMHD}
\end{document}